\documentclass[manuscript]{aastex631}
\submitjournal{PSJ}

\graphicspath{{./}}

\begin{document}

\title{Modeling the formation of Selk impact crater on Titan: Implications for Dragonfly}

\correspondingauthor{Shigeru Wakita}
\email{shigeru@mit.edu}

\author[0000-0002-3161-3454]{Shigeru Wakita}
\affiliation{Department of Earth, Atmospheric and Planetary Sciences, Massachusetts Institute of Technology, Cambridge, MA, USA}
\affiliation{Department of Earth, Atmospheric, and Planetary Sciences, Purdue University, West Lafayette, IN, USA}

\author[0000-0002-4267-093X]{Brandon C. Johnson}
\affiliation{Department of Earth, Atmospheric, and Planetary Sciences, Purdue University, West Lafayette, IN, USA}
\affiliation{Department of Physics and Astronomy, Purdue University, West Lafayette, IN, USA}

\author[0000-0003-3715-6407]{Jason M. Soderblom}
\affiliation{Department of Earth, Atmospheric and Planetary Sciences, Massachusetts Institute of Technology, Cambridge, MA, USA}

\author[0000-0002-7535-663X]{Jahnavi Shah}
\affiliation{Department of Earth Sciences, The University of Western Ontario, London, ON, Canada}

\author[0000-0003-3254-8348]{Catherine D. Neish}
\affiliation{Department of Earth Sciences, The University of Western Ontario, London, ON, Canada}
\affiliation{The Planetary Science Institute, Tucson, AZ, USA}

\author[0000-0002-1717-2226]{Jordan K. Steckloff} 
\affiliation{The Planetary Science Institute, Tucson, AZ, USA} 



\begin{abstract}
Selk crater is an $\sim$80 km diameter impact crater on the Saturnian icy satellite, Titan.
Melt pools associated with impact craters like Selk provide environments where liquid water and organics can mix and produce biomolecules like amino acids.
It is partly for this reason that the Selk region has been selected as the area that NASA’s Dragonfly mission will explore and address one of its primary goals: to search for biological signatures on Titan.
Here we simulate Selk-sized impact craters on Titan to better understand the formation of Selk and its melt pool. 
We consider several structures for the icy target material by changing the thickness of the methane clathrate layer, which has a substantial effect on the target thermal structure and crater formation.
Our numerical results show that a 4 km-diameter-impactor produces a Selk-sized crater when 5--15 km thick methane clathrate layers are considered.
We confirm the production of melt pools in these cases and find that the melt volumes are similar regardless of methane clathrate layer thickness.
The distribution of the melted material, however, is sensitive to the thickness of the methane clathrate layer. 
The melt pool appears as a torus-like shape with a few km depth in the case of 10--15 km thick methane clathrate layer, and as a shallower layer in the case of a 5 km thick clathrate layer.
Melt pools of this thickness may take tens of thousands of years to freeze, allowing more time for complex organics to form. 
 
\end{abstract}



\section{Introduction} \label{sec:intro}
NASA’s upcoming Dragonfly mission is scheduled to arrive at Titan in the 2030s to characterize its habitability, and search for potential chemical biosignatures \citep{Barnes:2021}.
One of the main science goals of the mission is to examine how liquid water may have mixed with organics on Titan’s surface, to produce molecules of prebiotic interest \citep[e.g.,][]{Neish:2010}.
To accomplish these goals, the Dragonfly spacecraft will target the region around the Selk impact crater for exploration \citep{Lorenz:2021}. 
Crater-forming impacts shock heat the target material,  generating melt pools of molten bedrock; like all large craters on Titan, Selk likely possessed a liquid water melt pool \citep{Neish:2018}.  
Such environments are conducive to the formation of prebiotic molecules such as amino acids \citep{Neish:2010}. 
Thus, investigating the production of impact melt pools is essential for understanding Titan's surface as a biological environment.

Selk crater is 84 $\pm 2$ km in diameter and $0.47^{+0.125}_{-0.105}$ km in depth \citep{Hedgepeth:2020}, with materials rich in water ice observed around the crater \citep{Soderblom:2010,Janssen:2016}, and organic-rich dunes beyond the crater rim \citep{Soderblom:2010,Lorenz:2021}.
Although impact melt formation has been examined previously for Titan \citep{Artemieva:2003,Artemieva:2005,Crosta:2021}, those works assumed that the target is water ice, ignoring the presence of methane clathrate in the substrate. 
Methane clathrate hydrate is stable at Titan's present-day surface temperature and pressure conditions and methane clathrate is expected to constitute a significant portion of Titan’s upper crust \citep{Choukroun:2010,Vu:2020}. 
Since methane clathrate is much stronger than water ice near the melting temperature of water ice \citep{Durham:2003}, the presence of methane clathrate in the target can affect the formation and evolution of Titan's impact craters. 
Indeed, previous work showed that a methane clathrate target results in slightly smaller craters than a pure water-ice target \citep{Wakita:2022}. 
As such, target materials would change the crater size for a given impactor size.   

The presence of methane clathrates, however, also has a strong effect on the target temperature at depth. 
Methane clathrate has a lower thermal conductivity (0.5 W/m/K) than water ice (2.2 W/m/K) \citep[at 263 K,][]{Sloan:2007}, and the presence of methane clathrate in the crust results in higher lithospheric thermal gradients than would exist for crusts made purely of water ice with the same heat flow \citep{Kalousova:2020}.
The target material weakens at these higher temperatures and becomes strengthless as the temperature approaches its melting temperature. 
Thus, the vertical temperature profile of the target significantly influences crater formation and acts as a primary control on crater morphology \citep{Bray:2014,Bjonnes:2022,Silber:2017}.
The effect of methane clathrate's thermal properties is also missing in previous works modeling the production of impact melts on Titan.

Here we explore the formation of a Selk-sized crater on Titan. 
We simulate impacts with different initial conditions, considering a top surface layer of methane clathrate, which is accompanied by a higher thermal gradient and increased strength. 
We explore how craters on Titan vary depending on the thickness of methane clathrate and the associated temperature profile.
We also examine the distribution of the impact induced melt.
By investigating the formation and evolution processes of Selk crater, with a particular focus on the formation and source of melt pools and the thermal effects of methane clathrates, we provide crucial information needed to understand one of the main targets of Dragonfly's scientific investigations.

\section{Methods} \label{sec:meth}
We perform crater-forming impact simulations on Titan using the iSALE-2D shock physics code.
The code has been developed to model planetary impacts and cratering \citep{Collins:2016, Wunnemann:2006}. 
It is based on the SALE hydrocode \citep{Amsden:1980}
and has been improved by including various equations of state (EOS) and strength models \citep{Collins:2004, Ivanov:1997, Melosh:1992}.
We assume that a spherical icy impactor hits a flat target with an impact velocity of 10.5 km/s, the average velocity of an impact on Titan \citep{Zahnle:2003}.
As we perform two-dimensional simulations in an axisymmetric coordinate system, all simulations explored here assume vertical impacts.
We consider spherical icy impactors ($D_{\rm imp}$) with diameters that ranges from 3 to 4 km.
iSALE-2D has a high-resolution zone and an extension zone.
We use a resolution of 50 m for the high-resolution zone that includes the impactor and the target down to a depth of 15 km and out to a radius of 70 km. 
The cell size in the extension zone is incrementally increasing by 2 \% from the previous cell up to a maximum cell size of 1000 m (20 times larger than the cell size in the high-resolution zone).
While we consider water ice as the composition of the impactor, we consider methane clathrate and water ice for the composition of the target.
Because shock physics EOS or Hugoniot data for methane clathrate is currently unavailable, we use an analytical equation of state (ANEOS) for water ice for both compositions (see below), and the strength model of methane clathrate developed in previous work \citep{Wakita:2022}. 
Material input parameters used in this work are shown in Table \ref{tab:input}.

We explore the sensitivity of Titan’s impacts to (1) the thickness of a methane clathrate layer overlying a solid water-ice layer and (2) the crustal temperature profile. 
Since the thermal conductivity of methane clathrate is much lower than that of water ice \citep{Sloan:2007}, 
the temperature gradient within the methane clathrate conductive lid is higher than for the water-ice case with the same heat flow as shown in \cite{Kalousova:2020}.
To explore the effect of the temperature gradient, we study a 10 km thick methane clathrate layer with various temperature gradients ranging from 5 K/km to 20 K/km.
When we assume that a conductive ice lid lays on top of a convective ice layer, 
the thickness of the conductive lid depends both on the temperature gradient and on the temperature of the convecting ice \citep{Silber:2017}.
Assuming a temperature of 255 K for the convecting ice \citep{Kalousova:2020}, we model the thickness of the conductive ice lid to be 32.4 km, 16.15 km, 10.75 km, and 8.06 km for temperature gradients of 5 K/km, 10 K/km, 15 K/km, and 20 K/km, respectively. 
We take the lower limit of 5 K/km, which is similar to the no methane clathrate case \citep[3 K/km]{Wakita:2022}, and the upper value from the thin methane clathrate case \citep[23 K/km]{Kalousova:2020}.
The water-ice layer beneath the 10 km thick methane clathrate layer is therefore in the conductive lid for all temperature gradients considered except for the 20 K/km case.
Using a constant temperature profile, however, can lead to errors in the heat flux. 
By definition, the heat flux must be the same on each side of the boundary between the methane clathrate layer and the water-ice layer. 
Because of the different thermal conductivity of the two materials, however, this requires different temperature gradients on each side of the boundary.  
Thus our simple temperature profile model that assumes the same temperature gradient at the boundary leads to an inaccurate heat flux. 
As such, we cannot trust the heat flux between the methane clathrate layer and the water-ice layer for these simple cases.
Nevertheless, our simplified models allow us to directly explore the effects of the temperature gradient on the crater morphology. 
We explore and discuss the effects of the temperature of the convective layer below (see \ref{sec:disc} Discussion).
Next, we consider temperature profiles for 5, 10, and 15 km thick methane clathrate layers, based on thermal modeling in \cite{Kalousova:2020};
these temperature profiles are likely to be more accurate because they enforce a continuous heat flux at the methane-clathrate--water-ice boundary.
Note that the temperature gradient in the conductive lid of methane clathrate is $\sim 23$, $\sim 15$, and $\sim 10$ K/km for 5, 10, and 15 km thick cases, respectively.
The temperature profiles and yield strengths of these seven models are shown in Figures \ref{fig:tmp_yld_255K} and \ref{fig:tmp_yld_KS}.

\begin{figure}
\plotone{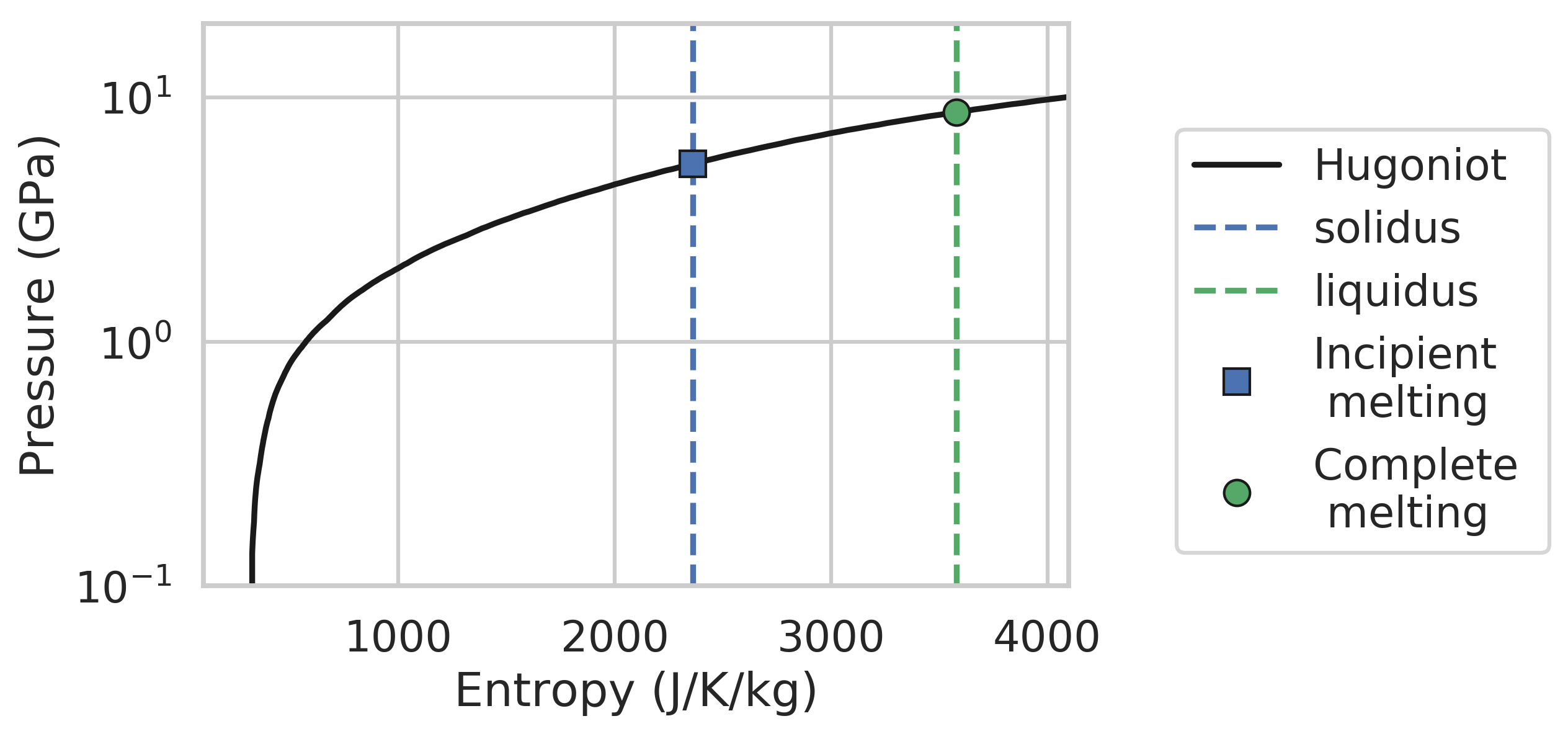}
\caption{
Hugoniot curve for water ice (black solid line). 
The blue dashed line depicts the solidus, and the green dashed line depicts the liquidus. 
We take a threshold pressure from the intersect between the solid and dashed lines (see symbols).
\label{fig:aneos}}
\end{figure}

To identify any material that becomes impact melt (i.e., liquid water), we rely on the peak pressure of tracer particles.
The Lagrangian tracer particles in iSALE-2D are placed in each cell at the start of the simulation prior to impact ($t$ = 0), and record the pressure at regular intervals throughout the impact process. 
Because they follow the parcel of material in which they are embedded, they are useful to estimate the status of melt.
Since ANEOS is a semi-analytical model derived from the first principles of thermodynamics, thermodynamical quantities of materials (e.g., temperature, pressure, and density) are obtained self-consistently from the Gibbs free energy \citep{Thompson:1990,Melosh:2007}.
In iSALE, we can use Tillotson EOS which is a simple analytical model using a simplified phase diagram. 
The Tillotson EOS, however, can not compute thermodynamical quantities consistently, especially at phase transitions.
Because ANEOS has the treatment of multiple phases, we use it in this work (see also Section \ref{sec:disc}).
Figure \ref{fig:aneos} shows the Hugoniot curve of water ice using ANEOS.
To determine the fate of material we compare the entropies of the solidus (2360 J/kg/K) and liquidus (3580 J/kg/K) to entropies and pressures along the Hugoniot curve. 
Incipient melting occurs when the material pressure exceeds the intersect of the solidus and Hugoniot curve (5.34 GPa). 
Complete melting occurs when its pressure exceeds the intersect of the liquidus and Hugoniot curve (8.69 GPa).
We adopt these thresholds to determine incipient and complete melting of materials during the impact process.
Note that these thresholds are similar to previous work that also uses ANEOS \citep{Artemieva:2003}, but differ from other work that uses thermochemical tables \citep{Pierazzo:1997}. 
While there is a more complex EOS for water ice (5--phase EOS), its estimated amount of complete melting is similar to that of ANEOS \citep{Kraus:2011}. 
Thus, the amount of melt predicted using the less complex ANEOS is reasonable. 

In this work, we consider the incipient and complete melt of tracer particles according to their peak pressure. 
We calculate their volumes based on their initial spacing and locations \citep[e.g.,][]{Johnson:2014}.
While we will discuss the volume of incipient and complete melt, we do not treat the volume in between them (i.e., partial melting tracer particles).

To account for melting of methane clathrate, we  apply the same approach.
Although the Hugoniot data of methane clathrate is unavailable, the melting curve of methane clathrate is given experimentally \citep{Sloan:2007}.
Methane clathrate has a similar phase curve to water ice in the high pressure environment \citep[$\gtrsim$ 1.5 GPa, ][]{Journaux:2020}. 
Since our threshold pressure of incipient melt (5.34 GPa) exceeds 1.5 GPa, we assume that methane clathrate follows the same criteria for dissociation as that of water ice. 
Moreover, methane clathrate may transform to water ice at lower pressure by releasing methane gas. 
The releasing methane gas would change the target material's volume, resulting in compression that would lead to an increase of internal energy and additional heating \citep{Melosh:1989}. 
The compaction of any pre-existing porosity in the target would also lead to similar compression and heating. 
As such, our melt estimates should be treated as lower limits. 

\section{Results} \label{sec:res}
\subsection{Effect of a temperature gradient in 10 km thick methane clathrate layer} \label{sec:10km}
We first consider the sensitivity of crater formation on the temperature gradients assuming a 10 km thick methane clathrate layer over water ice and a 4 km diameter icy impactor.
Figure \ref{fig:tmp_15Kb255K} illustrates the time series of crater formation with a temperature gradient of 15 K/km.
Soon after impact a transient crater (19 km in depth and 14 km in radius, Figure \ref{fig:tmp_15Kb255K}a) forms, and then collapses producing a central uplift (Figure \ref{fig:tmp_15Kb255K}b).
This central uplift then collapses, pushing warm material over the apparent crater rim (Figure \ref{fig:tmp_15Kb255K}c). 
Because the overflow of this warm material covers the rim, the rim is obscured when the crater is finally formed (Figure \ref{fig:tmp_15Kb255K}d, Figure \ref{fig:zoom_15Kb255K}, and supplemental movie Figure \ref{sup:mv}).
Because this overflow is likely exaggerated by the axisymmetry of our numerical simulations, 
the rim after the overflow is not representative of the crater rim. 
Similar behavior is observed in modeling crater formation on other icy satellites with higher temperature gradients such as Europa \citep{Silber:2017}. 
When such overflow occurs in these models, we must examine the crater diameter before this overflow covers the rim (i.e., the time of rim formation (Figure \ref{fig:tmp_15Kb255K}b)); following \cite{Silber:2017}, we take a rim location between the time that the transient crater collapses and the central uplift collapses. 
Figure \ref{fig:zoom_15Kb255K} depicts the surface profile before the overflow ($t$ = 400 s) and after the overflow ($t$ = 2000 s).
Although their rim locations are coincidentally near, the profile at the earlier time ($t$ = 400 s) is more pronounced than the later time ($t$ = 2000 s). 
As such, defining such a topographic height, which clearly exceeds the original surface as a rim,
we determine the crater diameter and depth before the overflow occurs \citep{Silber:2017}.
The crater depth is taken as the vertical distance between the rim height at this intermediate time step and the lowest location in the crater floor at the end of the simulation. 

The temperature gradient governs the diameter and depth of craters. 
The top four rows in Table \ref{tab:crater} show the diameters and depths of craters formed in a 10 km thick methane clathrate layer over water ice assuming different linear temperature gradients and a convective temperature of 255 K. 
In the case of the 5 K/km thermal gradient, the transient crater is shallow (16 km, see Figure \ref{fig:tmp_dimp4km_mc10}a) and remains entirely in a relatively high yield-strength region of the target. 
As such, craters formed in a 5 K/km target are relatively small and deep. 
For the higher thermal gradients (15 and 20 K/km cases), the transient crater is deeper ($\sim$  20 km in both cases, see Figure \ref{fig:tmp_dimp4km_mc10}) and reaches a lower yield-strength region that allows for greater uplift of the water-ice layer. 
Thus, craters formed in 15--20 K/km targets are wider and shallower and, because the yield strength at 20 km-depth is similar for both scenarios (Figure \ref{fig:tmp_yld_255K}), the depths are similar. 
The 10 K/km case is intermediate between these scenarios; most of the transient crater resides in the colder, higher-yield-strength material, with only the base of reaching the warmer and weaker material. 
As a result, craters formed in this target have diameters that are comparable to the 5 K/km case and depths that are similar to the 15 and 20 K/km cases (see Table \ref{tab:crater} and Figure \ref{fig:tmp_dimp4km_mc10}). 
It is worth noting that the crater depth can change by a few meters to a kilometer or more if we use a different convective temperature (see Section \ref{sec:disc}).
Note that the behavior of the rim relates to the crater size; the rim of the 20 K/km case behaves like the 15 K/km case (see Figure. \ref{fig:zoom_15Kb255K}).

\subsection{Effect of methane clathrate layer thickness} \label{sec:thick}
Next, we examine the dependence of crater formation on Titan to the thickness of the methane clathrate layer. 
Again assuming a 4 km diameter impactor, we consider 5, 10, and 15 km thick methane clathrate layers and their associated temperature profiles reported in \citet{Kalousova:2020} (see Section \ref{sec:meth} and Figure \ref{fig:tmp_yld_KS}). 
Our results show that the size of the transient crater is similar in all cases. 
This is because a 4 km-diameter-impactor with 10.5 km/s has enough energy to open a $\sim$ 20 km deep transient crater, which is deeper than the methane clathrate thickness considered in all of these cases.
Since the yield strength at the base of the transient crater ($\sim$ 20 km) is almost the same in the 10 and 15 km thickness cases (see Figure \ref{fig:tmp_yld_KS}), the resulting craters are of a similar depth crater (see Table \ref{tab:crater}). 
This same behavior is observed for 10--20 K/km temperature gradients (see the previous section). 
This suggests that the craters formed in the methane clathrate layer ($>$ 10 km) with a temperature gradient of 10--20 K/km would be comparable.
The water-ice layer is uplifted in these cases, resulting in a relatively shallow crater. 
The size of these craters is 85--96 km in diameter, yet only 1.1--1.7 km in depth.

\subsection{Effect of impactor size} \label{sec:impactor}
Next, we explore how smaller impactors affect crater formation and crater morphology on Titan.
We consider the same range of targets as in the previous section: a 5, 10, or 15 km thick methane clathrate layer  and the corresponding temperature profile \citep[Figure \ref{fig:tmp_yld_KS}, ][]{Kalousova:2020}.
Smaller impactors have less impact kinetic energy, and therefore open smaller transient craters; for both 3 and 3.5 km diameter impactors, the transient craters are less than 15 km in depth. 
We still observe uplifting of the water-ice layer in the case of 5 km and 10 km thick methane clathrate layers, but the uplift of water ice fails to occur if the methane clathrate layer is 15 km thick. 
This is because the transient crater forms entirely within the methane clathrate layer, which has a higher yield strength than the water-ice layer (see blue dashed line in Figure \ref{fig:tmp_yld_KS}).

The occurrence of water-ice uplift directly affects crater depth.
Figure \ref{fig:crater_KS} summarizes the diameter and depth of craters formed for the ranges of impactor sizes and methane clathrates layers we consider. 
Impact simulation studies typically assume uncertainties of 2 or 3 cells; we take 3 cells for the uncertainty on each side and depth of the crater (i.e., 6 cells). 
Although we plot these uncertainties as error bars, they are relatively small and lie almost within the symbols.
Since uplift of the water-ice layer occurs in the case of both 5 km and 10 km thick methane clathrate layers,
these crater depths are similar, regardless of the impactor size (see red and green symbols in \ref{fig:crater_KS}).
The depth of craters in a 15 km thick methane clathrate, however, is inversely proportional to the diameter of the impactor. 
Smaller impactors produce smaller and shallower transient craters and, in turn, also produce smaller and shallower final craters.
As discussed above, we determine the rim location before the crater collapse (e.g., Figure \ref{fig:zoom_15Kb255K}); for consistency, we do this regardless of the occurrence and absence of overflow.
 
\subsection{Production of the melt pool} \label{sec:melt}
Although similar sized craters can form under different structural conditions, the resulting distribution of heated materials varies significantly.
Figure \ref{fig:trp_dimp4kmKS} illustrates the distributions of tracer particles colored by their peak pressure for a 4 km-diameter-impactor into a target with a 5, 10, or 15 km methane clathrate layer and the corresponding temperature profile of \cite{Kalousova:2020}.
We color the tracer particles only if the peak pressure exceeds the incipient melting pressure (5.34 GPa, see Section \ref{sec:meth}). 
The black dotted lines indicate the material boundary; methane clathrate and water-ice distribute near the surface in cases of 5 km and 10 km thick methane clathrate layers, but methane clathrate dominates the 15 km thick cases.
For the case of a 5 km thick methane clathrate layer, the heated materials are located below the crater floor as a layered structure and distributed along the crater wall and rim (Figure \ref{fig:trp_dimp4kmKS}a).
On the other hand, the distribution of heated materials in the 10 km and 15 km thick layers are relatively complex (Figure \ref{fig:trp_dimp4kmKS}b, c). 
The folding of methane clathrate during the crater collapse creates torus-like structures of heated material (e.g., Figure \ref{fig:tmp_15Kb255K}).
Note that there is a gap between the surface (black dashed lines) and the tracer particles (i.e., Figure \ref{fig:trp_dimp4kmKS} a). 
When material collapses inward to the symmetry axis, a single tracer particle represents several cells of material. 
After this material collapses outward, those tracer particles corresponding to several cells deviate from the surface that is represented by the cell, resulting in potentially erroneous results at the boundary of the simulation; this may be the cause of the heated materials (colored tracer particles) being distributed along the axis in the 10 km thick case.
Figure \ref{fig:trp0_dimp4kmKS} illustrates the provenance plots of melt with the same impact conditions as in Figure \ref{fig:trp_dimp4kmKS}. 
This clearly shows that the heated material originates only from the methane clathrate layer in the 10 km and 15 km thick layers. 
On the other hand, the water-ice layer also contributes to the heated material for the 5 km thick layer case.
Overall, the thickness of methane clathrate affects the distribution of heated material and its origin.

Our results confirm that impact induced heat can produce melt pools in Selk-sized craters on Titan and demonstrate that such melt occurs in the presence of a methane clathrate layer. 
Figure \ref{fig:melt_dimp4kmKS} represents the incipient and complete melt volume corresponding to the horizontal direction with the same impact conditions as in Figure \ref{fig:trp_dimp4kmKS}. 
Note that we calculate the volume of each cell using the initial location of the tracer particles (see Section \ref{sec:meth}).
For a given impactor diameter, the total incipient volume of melt is similar regardless of the initial target structure (thickness of methane clathrate): A 4 km-diameter-impactor produces $\sim$ 360--400 km$^3$ of melt, a 3.5 km-diameter-impactor produces $\sim$ 240--260 km$^3$, and a 3 km-diameter-impactor produces $\sim$ 150--160 km$^3$, respectively. 
\footnote{We use the thresholds of peak pressure for incipient melt as 5.34 GPa and that for complete melt as 8.69 GPa (see Section \ref{sec:meth}).}
The total volume of complete melt is also similar (190--220 km$^3$, 130--140 km$^3$, 80--90 km$^3$ for 4, 3.5 and 3 km-diameter impactors, respectively).
We note that higher impact velocities will produce larger total melt volumes \citep{Pierazzo:1997,Artemieva:2005}.
A 3 km-diameter-impactor with a higher impact velocity would produce a similar amount of melt as a 4 km-diameter-impactor at 10.5 km/s. 
And finally, as we ignore melted ejecta excavated from the crater, these total volumes include only material within the crater.
As Figure \ref{fig:trp0_dimp4kmKS} shows, the total volume of (incipient/complete) melt is almost completely independent of the methane clathrate thickness.
Nevertheless, as mentioned above, the spatial distribution of melt is affected, 
with melt being concentrated within 20 km of the crater center (see Figures \ref{fig:trp_dimp4kmKS}b, c and Figure \ref{fig:melt_dimp4kmKS}b, c) if the clathrate layer is $\geq$ 10 km thick, 
and being broadly distributed over the crater floor, wall, and rim, if the clathrate later is 5 km thick (see Figures \ref{fig:trp_dimp4kmKS}a and \ref{fig:melt_dimp4kmKS}a). 
These differences in the distribution of the melt will dramatically affect freezing time, because the structure (e.g., thickness) governs the thermal evolution of the melted material.
Additionally, the distribution of melt (i.e., liquid water) will influence its ability to mix with the surface material, including possible organic materials, with the wide distribution of the 5 km clathrate layer case (Figure \ref{fig:trp_dimp4kmKS}a) having the greatest potential for such mixing.
This implies that the chemistry of the melt pool in a 5 km thick case might be different from other cases. 
Therefore, even if the total melt volume is similar, the structure and chemistry of the melt pool may differ depending on the thickness of methane clathrate. 

\section{Discussion} \label{sec:disc}
\begin{figure}
\plotone{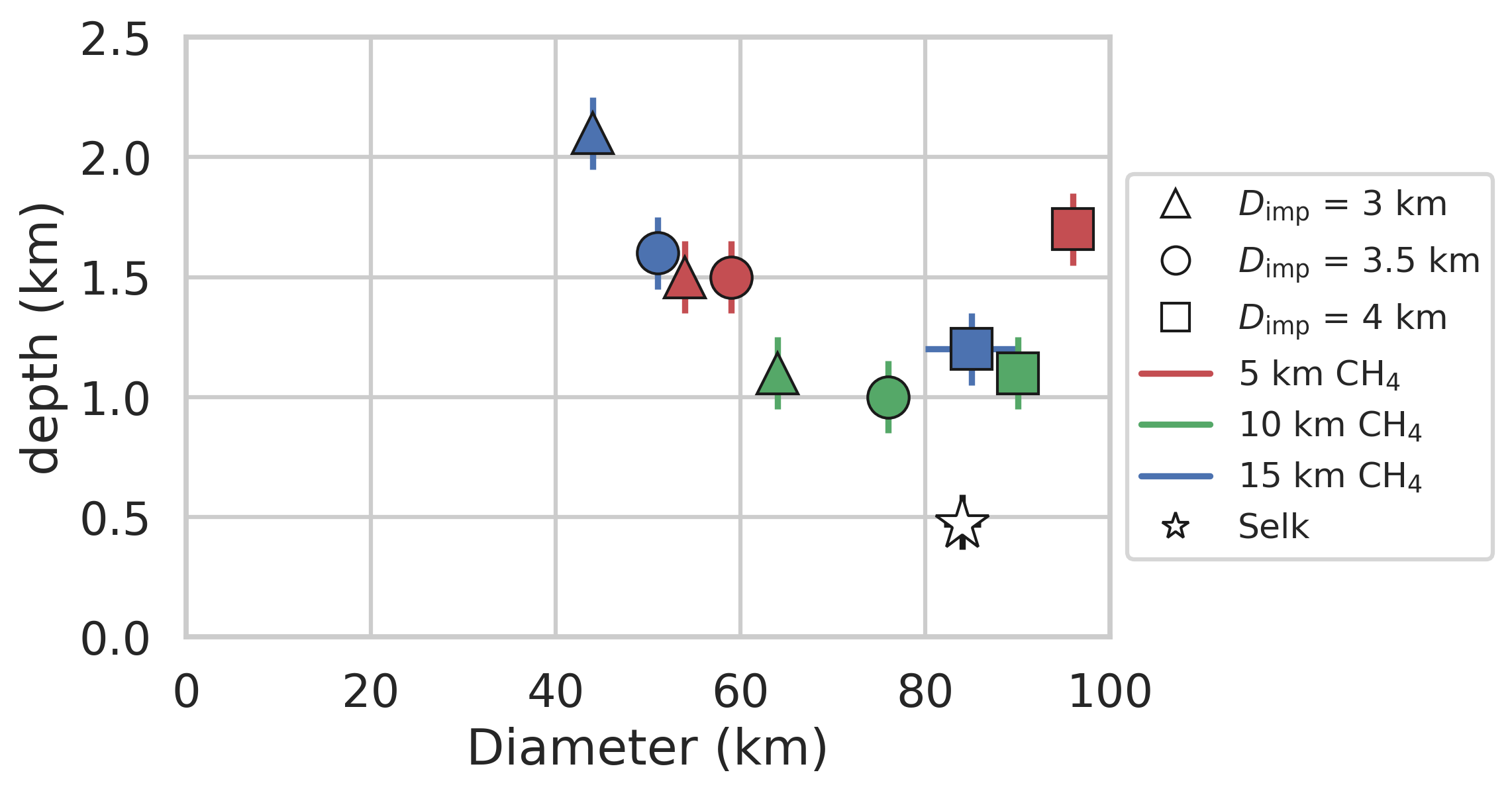}
\caption{
Crater depth as a function of crater diameter. 
These craters are formed by 3--4 km-diameter-impactors on a 5--15 km thick methane clathrate layer with the corresponding temperature profiles from \citet{Kalousova:2020} (see legend).
While the error bars correspond to the uncertainty of six cells, they are almost within the symbols.
Note that we also added the horizontal error bar in the case of the very ambiguous rim (see blue square symbol, 15 km thick methane clathrate with 4 km-diameter-impactor), which makes it hard for us to define the crater rim.
Selk crater is also plotted as a star with error bars \citep[again almost within the symbol,][]{Hedgepeth:2020}.
\label{fig:crater_KS}}
\end{figure}

\begin{figure}
\plotone{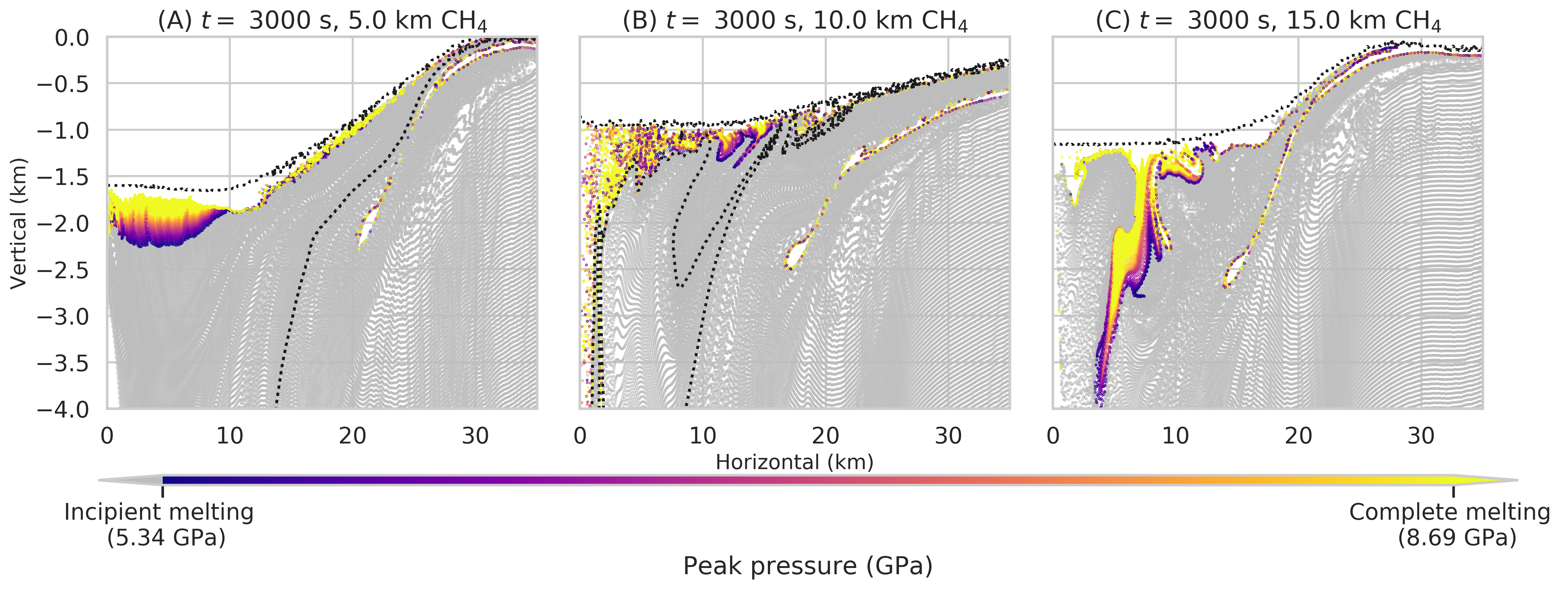}
\caption{
Distribution of tracer particles, colored by their peak pressure.
Unmelted particles are indicated by gray color, and the black dashed lines depict material boundaries.
These craters are formed by 4 km-diameter-impactors with the temperature profile of \cite{Kalousova:2020} (see square symbols in Figure \ref{fig:crater_KS}). 
\label{fig:trp_dimp4kmKS}}
\end{figure}

\begin{figure}
\plotone{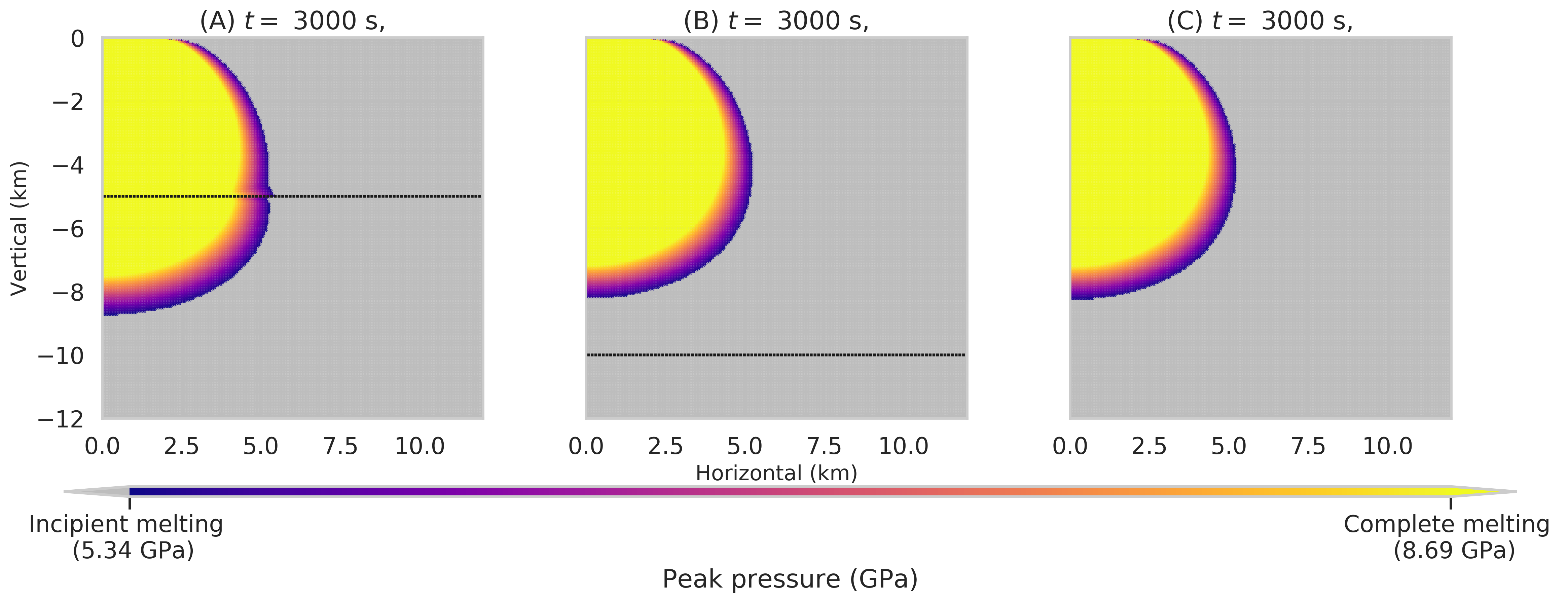}
\caption{
Same data in Figure \ref{fig:trp_dimp4kmKS}, but showing the provenance plots. 
Tracer particles plotted at their original preimpact locations. 
\label{fig:trp0_dimp4kmKS}}
\end{figure}

\begin{figure}
\plotone{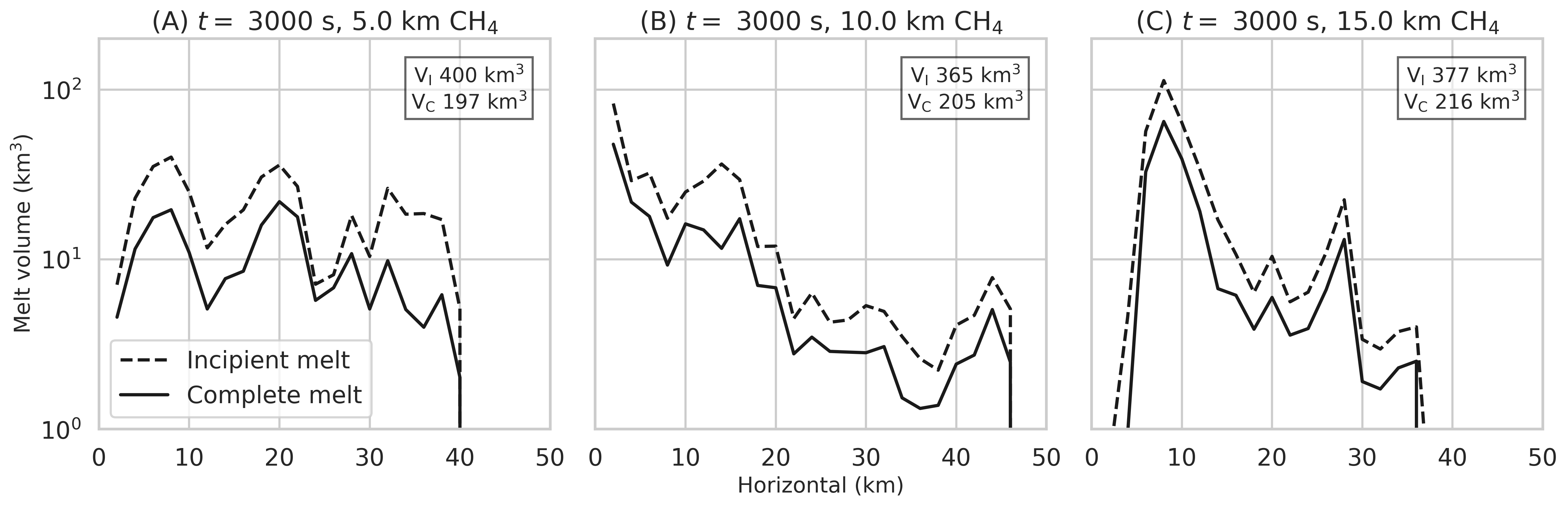}
\caption{
Distribution of melt volume as a function of the radial distance. 
These are products of 4 km-diameter-impactors with the temperature profile of \cite{Kalousova:2020} (see square symbols in Figure \ref{fig:crater_KS}). 
The dashed line depicts the incipient melt (5.34 GPa), and solid line depicts the complete melt (8.69 GPa). 
Note that we take the bin size in the horizontal direction as 2 km.
\text{
The box in the upper right indicates the total volume of incipient melt ($V_I$) and complete melt ($V_C$). 
}
\label{fig:melt_dimp4kmKS}}
\end{figure}

Our results indicate that impactors a few km in diameter striking a warm target with a surface layer of methane clathrate that is 5--15 km thick can produce Selk-like craters. 
Furthermore, we find that impacts into targets with higher temperature gradients produce craters that are shallower and wider than those formed in targets with lower temperature gradients (Table \ref{tab:crater}).
A 4 km diameter impactor striking a warm surface with a 10 km or 15 km thick methane clathrate layer forms the most analogous Selk-sized crater (see square green and blue symbols in Figure \ref{fig:crater_KS}), but a 10 km thick methane clathrate layer with $D_{\rm imp}$ = 3.5 km can also be a candidate (see green circle in Figure \ref{fig:crater_KS}). 
These results suggest that a target with a 10--15 km methane clathrate layer produces similar size craters if the impactor is large enough to make a transient crater that is $>$ 15 km deep. 
This is because the yield strength below 15 km depth is similar for such targets (Figure \ref{fig:tmp_yld_KS}),
because of the reasonable temperature profiles considering the lower thermal conductivity of methane clathrate.
On the other hand, the yield strength below 15 km depth in the case of the 5 km methane clathrate layer is higher than other cases (Figure \ref{fig:tmp_yld_KS}), thus the resultant crater differs.
This similarity in size only holds when an impactor forms a transient crater deeper than 15 km, which we only see in the 4 km-diameter-impactor cases.

\begin{figure}
\plotone{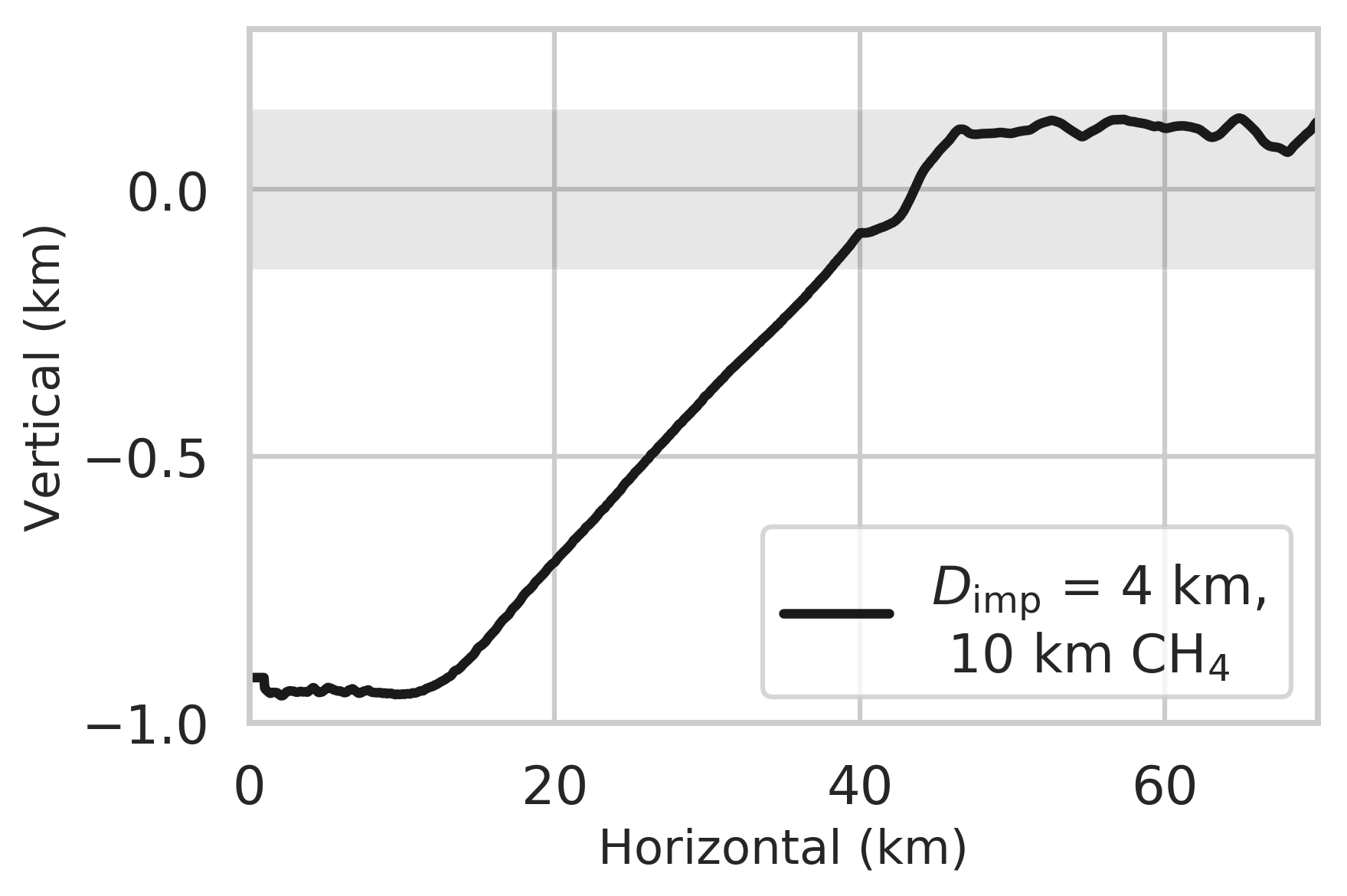}
\caption{
The averaged surface profile for our best-case simulation for producing a Selk-like crater. 
The impact condition is a 4 km diameter impactor hitting a target with a 10 km thick methane clathrate layer and the associated temperature profile from \citet{Kalousova:2020}.
The thick line denotes the averaged surface profile with a bin size of 2 km (40 cells) using the two surface profiles at different time steps.
We take $t$ = 400 s as the profile from 45 km and beyond, as this best represents the morphology of the rim prior to overflow (this is the time at which we determine the rim location ($\sim 45$ km)). 
For the profile from the center to 40 km, we use the profile at $t$ = 3000 s as the crater interior has stabilized by this time; this is demonstrated by the relatively flat crater floor compared to $t$ = 2000 s (their depths are almost identical).
For the profile of the intermediate region (40--45 km), we interpolate those profiles.
The shaded region denotes three cells from the original surface (six cells in total height).
\label{fig:selk}}
\end{figure}

Post-impact fluvial and aeolian erosion may explain the difference in depth between the model outputs and the measurements of Selk.
For the $\sim$1 km deep craters produced in a warm methane clathrate layer, we would require $\sim$0.5 km of infilling to reach the reported depth of Selk crater of $0.47^{+0.125}_{-0.105}$ km \citep{Hedgepeth:2020}. 
Although we can resolve the rim height with 2--3 cells in our model (e.g., $t$ = 400 s in Fig.  \ref{fig:zoom_15Kb255K}), here we attempt to compare it with the observed rim height. 
For our best case for Selk, we here consider a 4 km-diameter-impactor hitting a target with a 10 km thick methane clathrate layer the associated temperature profile from \citet{Kalousova:2020} (Figs. \ref{fig:crater_KS}, \ref{fig:selk} and Table \ref{tab:crater}).
Our rim height would be $\sim$0.15 km and lower than the Selk's rim height 
\citep[$0.280^{+0.050}_{-0.054}$ km, ][]{Hedgepeth:2020}. 
While we may underestimate the rim height, the depth from the original surface of $\sim$0.95 km would be deeper than the observed terrain depth of $0.190^{+0.155}_{-0.130}$ km \citep{Hedgepeth:2020}. 
In this case, the required amount of infilling would be $\sim$0.75 km which is slightly more than our estimate from the crater depth.
This level of infill is consistent with moderate amounts of fluvial erosion on Titan \citep{Neish:2016}. 
However, the sharp crater rim of Selk observed in the SARTopo data \citep{Hedgepeth:2020,Lorenz:2021a} suggests that infilling by sand might be a more reasonable explanation, given that fluvial erosion quickly flattens crater rims \citep{Neish:2016}. 
A thick deposit of sand in the crater can also insulate the crater floor, such that moderate amounts of viscous relaxation can further reduce the depth \citep{Schurmeier:2018}. 
Indeed, evidence of sand is visible in the Selk crater cavity in Cassini RADAR, VIMS, and ISS data \citep{Lorenz:2021}.
The interpretation that Selk might have been deeper when it formed ($\sim$ 1 km) is further supported by radarclinometry results that find variations in the height of the rim of Selk \citep{Bonnefoy:2022}. 
Further information from Dragonfly will help us to constrain the processes that have occurred at Selk.

It is possible that the ice shell structure at the time Selk formed might be different from our model, which would change the morphology of the crater. 
For example, if the heat flux from Titan's core was higher when Selk formed, the temperature gradient would be higher.
Since the icy shell moves more readily if the temperature gradient is higher, a higher thermal gradient would make the crater shallower than the results presented here \citep[e.g.,][]{Bjonnes:2022}.
Alternatively, the methane clathrate layer may have contained impurities when Selk formed, such as volatiles or liquids, which would weaken it. 
For example, if the crust contained more liquid methane and ethane than could diffuse into the ice shell to form clathrates \citep{Choukroun:2010,Vu:2020}, the crust would form a fluid-saturated layer.
Fluid-saturated layers behave like weak sediment, limiting the topography of craters, making them shallower than observed here \citep{Wakita:2022}.

\begin{figure}
\plotone{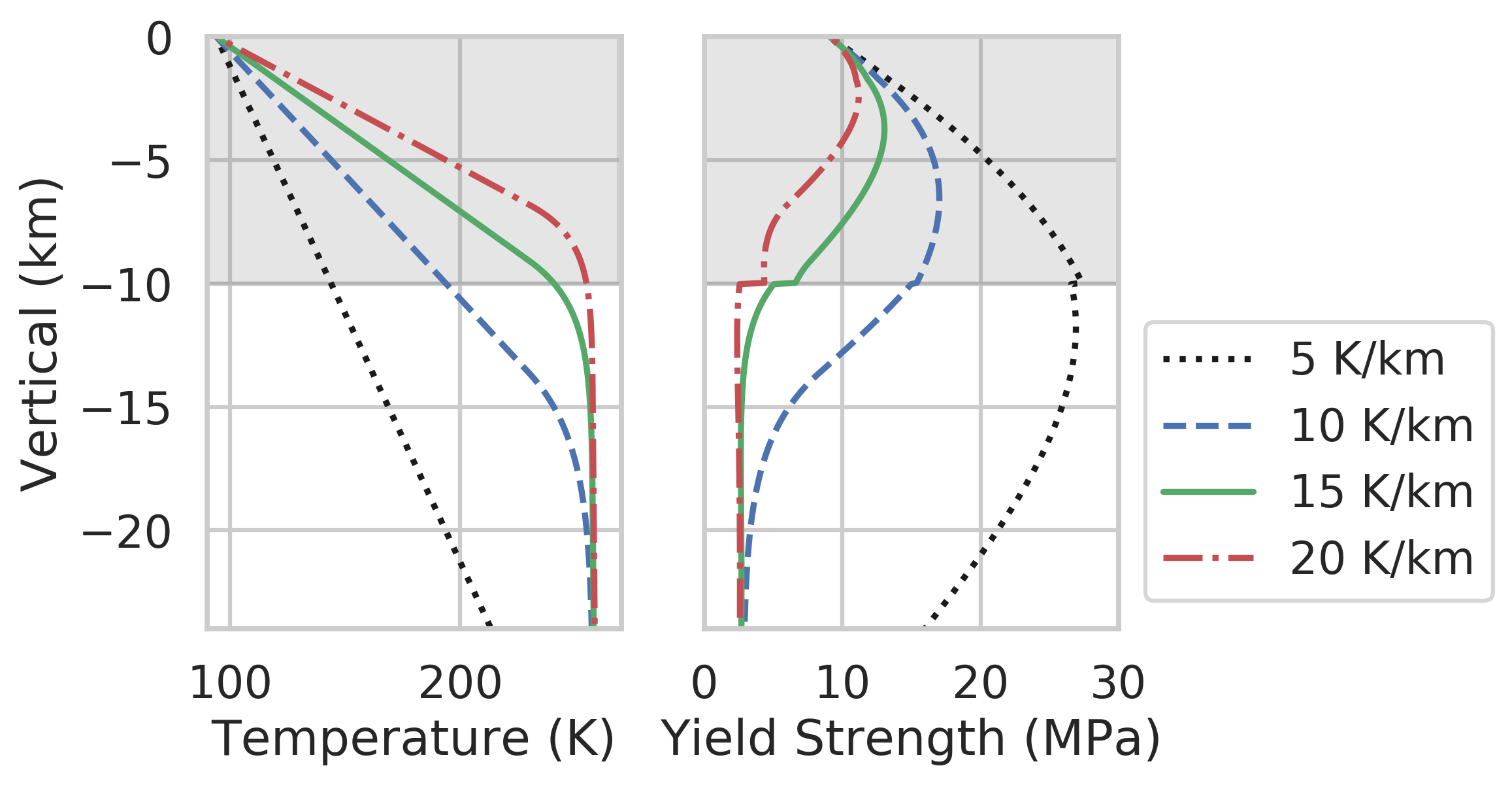}
\caption{
Temperature profile (left) and yield strength (right) plotted versus ice shell depth.
Each line represents a different temperature gradient in the conductive ice layer (see legend).
Note that we plot cases which have the convective temperature of 255 K.
The shaded region illustrates the 10 km thick methane clathrate layer.
Note that the jump in the yield strength indicates the boundary between methane clathrate and water-ice.
\label{fig:tmp_yld_255K}}
\end{figure}

\begin{figure}
\plotone{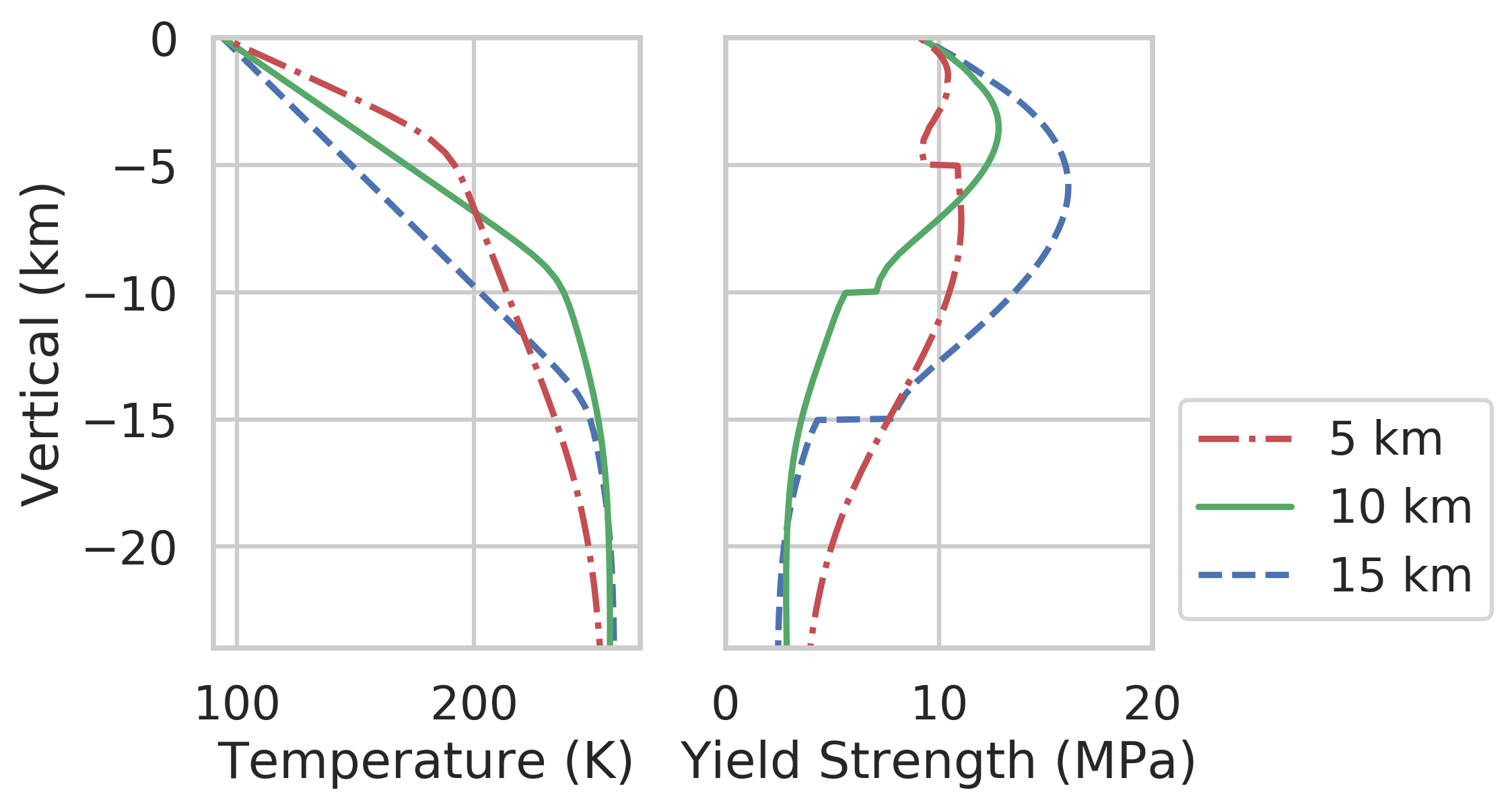}
\caption{
Similar to Figure \ref{fig:tmp_yld_255K}, but for the various methane clathrate thicknesses.
Each line represents a different methane clathrate thickness case with a corresponding temperature profile following \cite{Kalousova:2020}.
Note that the temperature gradient in the conductive lid of methane clathrate is $\sim 23$, $\sim 15$, and $\sim 10$ K/km for 5, 10, and 15 km thick cases, respectively.
Since the thermal softening parameter of methane clathrate (i.e., the dependence on temperature) differs from that of water-ice (see Table \ref{tab:input}), 
the yield strength of water-ice is slightly higher than that of methane clathrate at the boundary of the 5 km case (see red dash-dotted line).
\label{fig:tmp_yld_KS}}
\end{figure}

Moreover, the choice of the temperature in the warm convective layer (i.e., convective temperature) significantly affects crater depth.
While we took the convective temperature of 255 K for simple temperature profile cases (Section \ref{sec:10km} and Figure \ref{fig:tmp_yld_255K}), thermal modeling suggests plausible values that range from 254 K to 258 K \citep[Section \ref{sec:thick}, Figure \ref{fig:tmp_yld_KS}, ][]{Kalousova:2020}.
Since the temperature governs the yield strength, higher convective temperatures lead to weaker ice layers.
Our results of changing the convective temperature profile to 250 K and 260 K are shown in Table \ref{tab:crater}. 
Although there is almost no effect on depth for different convective temperatures in the 5 K/km cases, the depth of craters changes in 10--20 K/km cases (Table \ref{tab:crater}). 
When we use the lower convective temperature of 250K, the craters are deeper by a few hundred meters.
When we increase the convective temperature to 260 K, the craters are shallower by a few hundred meters to more than a km, depending on the temperature gradient.
Additionally, in case of temperature gradients of 15 and 20 K/km, the impacts produce little topography, and likely would be unobservable from orbital images (see bottom two rows in Table \ref{tab:crater}).
The effect of the temperature gradient and the convective temperature needs to be explored more for a variety of craters on Titan.

Although we assume a fixed impact velocity of 10.5 km/s in this study, different impact velocities are also possible. 
For different impactor velocities, we may use the crater scaling laws \citep{Melosh:1989}: Lower impact velocities at a given size of impactor forms a smaller crater. 
However, our choice of impact velocity may also affect the assumed acoustic fluidization parameters.
Acoustic fluidization models the weakening mechanism of the target during crater-forming impacts \citep{Melosh:1979}.
The simple block model is widely used to describe this, assuming a block size of oscillation \citep{Ivanov:1997}.
\citet{Silber:2017a} explored the effects of scaling block size; one is based on the impactor size \citep{Wunnemann:2003} and the other is based on transient crater size \citep[i.e., $\pi$ scaling,][]{Ivanov:2002a}.
They found that craters can be the same size but differ in morphology determined by the choice of scaling. 
They proposed the scaling of block size by the impactor size is preferable to explain observed lunar craters, especially in a range of 15--20 km diameter (i.e., simple-to-complex craters). 
We consider the parameters for the acoustic fluidization by using the block size scaling based on the impactor size, following the previous work of icy craters on Ganymede \citep{Bray:2014}.
Although their impact velocity of 15 km/s is higher than ours of 10.5 km/s, we assume that the parameters for acoustic fluidization are independent of impact velocity. 
We know, however, that the parameters for acoustic fluidization increase with the impact velocity when the parameters are based on transient craters \citep{Silber:2017a}. 
As such, it is possible that parameters based on the impactor size might also be a function of impact velocity.
The sensitivity of the crater diameter to acoustic fluidization parameters is minor \citep{Bray:2014}, but the crater depth is sensitive to these parameters. 
Using the same acoustic fluidization parameters in this study (i.e., independent of impact velocity), we have tested a 3.5 km-diameter-impactor with a velocity of 15 km/s hitting a 10 km thick methane clathrate layer with the thermal profile as in Figure \ref{fig:tmp_yld_KS}.
As expected, the resulting crater is larger in diameter (94 km) than the crater formed by the same-size impactor hitting at 10.5 km/s (76 km) and in fact is close to the diameter of crater formed by a 4 km-diameter-impactor traveling at 10.5 km/s (90 km, see Table \ref{tab:crater}).
Interestingly, a crater formed by this higher-velocity impactor is not substantially deeper (1.2 vs 1--1.2 km, Table \ref{tab:crater}) than a crater formed by similarly sized impactors impacting with a lower velocity.
This is because the uplift of water-ice layer causes shallow craters in all cases (see Section \ref{sec:thick}). 
Thus, unlike other bodies, impactor velocity may exhibit a greater control on the diameter of Titan's impact craters than on the depth of its craters. 

There are other numerical considerations in our simulation methodology that may also influence the crater depth. 
In this work, we used the ANEOS for water ice, but there are other EOSs for water ice.
We tested the Tillotson EOS for water ice \citep{Tillotson:1962,Ivanov:2002} and found that runs with the Tillotson EOS tend to form slightly shallower craters compared to ANEOS.
While ANEOS includes the phase transition (i.e., melting of water-ice), the Tillotson EOS does not. 
As a result, Tillotson overestimates the amount of heated (and melted) material and makes the material move more readily. 
To account for the melt pool formed by the crater forming impacts, our results using ANEOS are more accurate.

The production of a melt pool at Selk is extremely important for aspects of astrobiology on Titan. 
Organic molecules can react with the liquid water to form prebiotic molecules over timescales of years \citep{Neish:2010}, but the longer the liquid water remains before freezing, the more time the system has to produce more complex molecules. 
Thus, we briefly discuss the lifetimes of the melt pool here. 
A crater formed in a 15 km thick methane clathrate layer has a melt pool 8--10 km wide and 2--3 km deep (see Figures \ref{fig:trp_dimp4kmKS} and \ref{fig:melt_dimp4kmKS}).
It would take a torus shape, but we can estimate a freezing timescale from the depth of the deposit, assuming it is losing its heat through thermal conduction to the surface. 
For 1 km and 4 km thick deposits of liquid water on Titan, \citep{Neish:2006} estimated freezing timescales are 5000 yrs and 90,000 yrs, respectively. 
Therefore, these deposits may last for tens of thousands of years before freezing completely. 
These timescales are much longer than those calculated by \citet{Hedgepeth:2022} for melt pool a few hundred meters thick, who found that it takes just a few hundred years to freeze such melt pools. 
These estimates for melt pool depth came from previous modeling work, which assumed impact into cold water ice. 
Our results show that the presence of both methane clathrate and a warmer temperature profile result in much deeper melt pools with significantly extended lifetimes. 
This would increase the likelihood of the formation of complex organics on Titan.
These molecules would then concentrate within the melt sheet at different depths \citep{Hedgepeth:2022}. 
If these melt sheets are later eroded by fluvial processes, exposing these layers, Dragonfly may be able to detect the prebiotic species formed there \citep{Neish:2018}.
However, subsequent simulations are necessary to better constrain the lifetime and concentration of complex organic (carbon-hydrogen-nitrogen) molecules in the torus-shaped melt pool.

Note that since our simulations are axisymmetric, we are unable to reproduce some features of Selk crater. 
For example, Selk exhibits a polygonal shaped rim, which is most pronounced at its north-east \citep[i.e., "corner",][]{Soderblom:2010}.
Such polygonal-shaped structure can be seen on other planets and moons \citep[e.g., Meteor Crater, Arizona, USA,][]{Shoemaker:1963}. 
One hypothesis for their formation involves the pre-existence of faults that are exploited during the crater formation process \citep{Melosh:1989}.
Because we perform 2D impacts in an axisymmetric coordinates systems, all resultant craters become circular shapes. 
3D simulations capable of producing more complex shape craters would be necessary to model the polygonal shape of Selk crater.
Note that the oblique impacts in the 3D simulations may enlarge the melt volume within the methane clathrate layer \citep{Wakita:2019a,Wakita:2022b}. 
Nevertheless, their change in the topography might be modest \citep{Davison:2022}.
As we focus on the crater size (diameter and depth), the fate of impact ejecta is also beyond the scope of our work.
Future work on impact ejecta may elucidate the origin of the several hundred kilometer long landform east-southeast of Selk \citep["bench",][]{Soderblom:2010} that has been hypothesized to be a fluidized-ejecta flow, similar to the impact melt flows that are found around craters on Venus \citep{Phillips:1991,Schaber:1992} and the Moon \citep{Neish:2014a}.

\section{Conclusions} \label{sec:conc}
We performed impact simulations where an icy impactor hits a methane clathrate crust over water-ice on Titan.
The lower thermal conductivity of methane clathrate leads to a higher temperature profile in the methane clathrate ice shell than in a pure water-ice target. 
If a 5--15 km thick warm methane clathrate layer exists on Titan’s surface, a 4 km-diameter-impactor would produce a Selk-sized crater. 
While the resultant craters are shallower than the lower temperature gradient case, they are still slightly deeper than observed for Selk. 
Post-impact erosional processes such as fluvial erosion and aeolian infill likely explain the discrepancy in the crater depth; measurements by the DragonCam instrument on Dragonfly will help us to determine if such post-impact modification can explain this discrepancy or if additional factors must be considered in the crater formation process \citep{Barnes:2021}.

We also confirmed the production of large amounts of impact melt beneath Selk-sized craters, particularly if there is an insulating methane clathrate layer that results in a warmer crustal temperature profile.
While the total volume of the melt pool is similar for a given impactor size (360--400 km$^3$ for 4 km-diameter impactors) for all methane clathrate layer thicknesses we consider, the methane clathrate layer thickness affects the distribution of the melt.
The torus-like melt pool produced in a 10--15 km thick methane clathrate layer would take a longer time to freeze --- perhaps as long as tens of thousands of years --- than a shallower melt pool produced in a target with a 5 km thick methane clathrate surface layer.
If a thick layer of methane clathrate existed on Titan’s surface when Selk was formed, there was ample time for complex prebiotic chemistry to occur in the melt pool, increasing the likelihood of producing molecules of biological interest. 
Measurements by the Dragonfly Mass Spectrometer instrument on Dragonfly (DraMS) will be able to confirm these predictions \citep{Grubisic:2021}.

\begin{acknowledgments}
We gratefully acknowledge the developers of iSALE-2D, including Gareth Collins, Kai W\"{u}nnemann, Dirk Elbeshausen, Tom Davison, Boris Ivanov, and Jay Melosh (\url{https://isale-code.github.io/}).
We also thank Tom Davison, the developer of pySALEPlot tool, which helped us make some plots in this work.
This research was supported in part through computational resources provided by Information Technology at Purdue, West Lafayette, Indiana.
This work was supported by Cassini Data Analysis Program grant 80NSSC20K0382.
We thank the two reviewers for their helpful feedback.
The model inputs and outputs are available on Harvard Dataverse upon the acceptance.
\end{acknowledgments}





\bibliography{./Selk_forming_impacts.bbl}{}
\bibliographystyle{aasjournal}

\begin{deluxetable*}{lcc}
\tablecaption{iSALE input parameters \label{tab:input}}
\tablewidth{0pt}
\tablehead{
\colhead{Description} & \colhead{Methane clathrate} & \colhead{Water ice$^a$} }
\startdata
Equation of state & \multicolumn2c{ANEOS, H$_2$O}  \\
Thermal softening parameter & 0.8$^b$ & 1.2 \\
Cohesion, undamaged (MPa) & 10 & 10 \\
Cohesion, damaged (MPa) & 0.01 & 0.01 \\
Frictional coefficient, undamaged & 2.0 & 2.0 \\
Frictional coefficient, damaged & 0.6 & 0.6 \\
Limiting strength, undamaged (GPa) & 2.2$^c$ & 0.11 \\
Limiting strength, damaged (GPa) & 2.2$^c$ & 0.11 \\
\enddata
\tablecomments{Parameters are the same as that of water ice$^a$ unless otherwise mentioned.}
\tablenotetext{a}{\cite{Bray:2014, Silber:2017}}
\tablenotetext{b}{\cite{Wakita:2022}}
\tablenotetext{c}{Representing 20 times stronger methane clathrate than water ice \citep{Durham:2003}}
\end{deluxetable*}

\begin{deluxetable*}{cccccc}
\tablecaption{Crater diameter and depth \label{tab:crater}}
\tablewidth{0pt}
\tablehead{
\colhead{Impactor} & \colhead{Crater} & \colhead{Crater} & \colhead{Methane clathrate} 
& \colhead{Temperature} & \colhead{Convective} \\
\colhead{diameter [km]} & \colhead{diameter [km]} & \colhead{depth [km]} & \colhead {thickness [km]} 
& \colhead{gradient [K/km]} & \colhead{ temperature [K]} 
}
\startdata
4 & 61 & 3.2 & 10 & 5 & 255 \\
4 & 60 & 1.4 & 10 & 10 & 255 \\
4 & 90 & 1.0 & 10 & 15 & 255 \\
4 & 97 & 1.2 & 10 & 20 & 255 \\
4 & 96 & 1.7 & 5 & \multicolumn{2}{c}{\cite{Kalousova:2020}$^a$} \\
4 & 90 & 1.1 & 10 & \multicolumn{2}{c}{\cite{Kalousova:2020}$^a$} \\
4 & 85 & 1.2 & 15 & \multicolumn{2}{c}{\cite{Kalousova:2020}$^a$} \\
3.5 & 59 & 1.5 & 5 & \multicolumn{2}{c}{\cite{Kalousova:2020}$^a$} \\
3.5 & 76 & 1.0 & 10 & \multicolumn{2}{c}{\cite{Kalousova:2020}$^a$} \\
3.5 & 51 & 1.6 & 15 & \multicolumn{2}{c}{\cite{Kalousova:2020}$^a$} \\
3 & 54 & 1.5 & 5 & \multicolumn{2}{c}{\cite{Kalousova:2020}$^a$} \\
3 & 64 & 1.1 & 10 & \multicolumn{2}{c}{\cite{Kalousova:2020}$^a$} \\
3 & 44 & 2.1 & 15 & \multicolumn{2}{c}{\cite{Kalousova:2020}$^a$} \\
4 & 62 & 3.1 & 10 & 5 & 250 \\
4 & 61 & 1.7 & 10 & 10 & 250 \\
4 & 89 & 1.3 & 10 & 15 & 250 \\
4 & 90 & 1.2 & 10 & 20 & 250 \\
4 & 62 & 3.2 & 10 & 5 & 260 \\
4 & 56 & 0.5 & 10 & 10 & 260 \\
4 & NA$^b$ & NA$^b$ & 10 & 15 & 260 \\
4 & NA$^b$ & NA$^b$ & 10 & 20 & 260 \\
\enddata
\tablenotetext{a}{We use the temperature profile in \cite{Kalousova:2020}. See also Figure \ref{fig:tmp_yld_KS}.}
\tablenotetext{b}{Almost flat with little topography.}
\tablecomments{All craters are given by vertical impacts with 10.5 km/s.}
\end{deluxetable*}

\begin{figure}
\epsscale{0.6}
\plotone{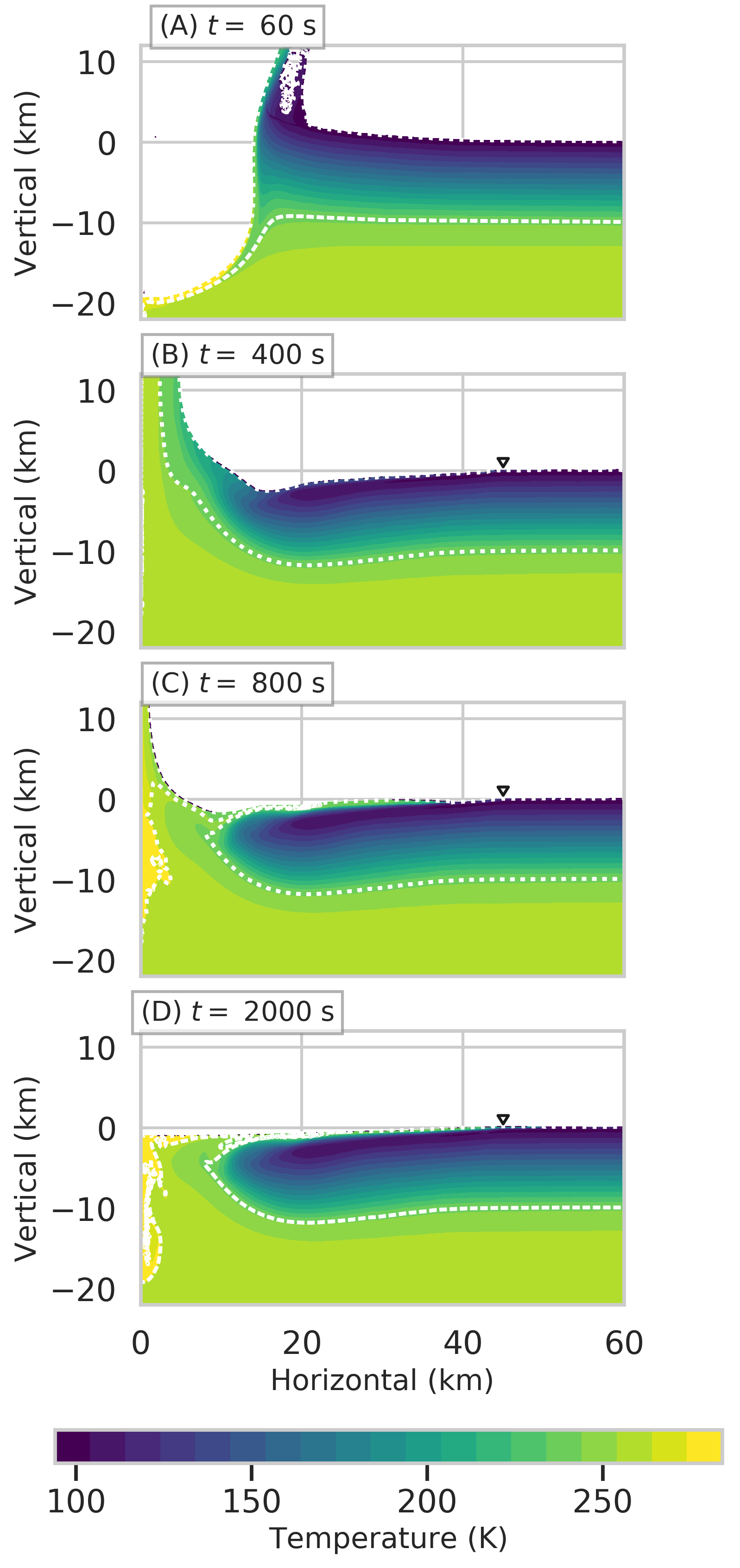}
\caption{
Time series of crater formation for a 4 km-diameter-impactor hitting a 10 km thick methane clathrate layer over water ice with a temperature gradient of 15 K/km.
Color indicates the temperature in Kelvin, and the white dotted lines represent the material boundary. 
Arrows indicate the rim location (see also Figure \ref{fig:zoom_15Kb255K}).
Supplemental movie Figure \ref{sup:mv} covers the time sequences that are not shown here.
\label{fig:tmp_15Kb255K}}
\end{figure}

\begin{figure}
\plotone{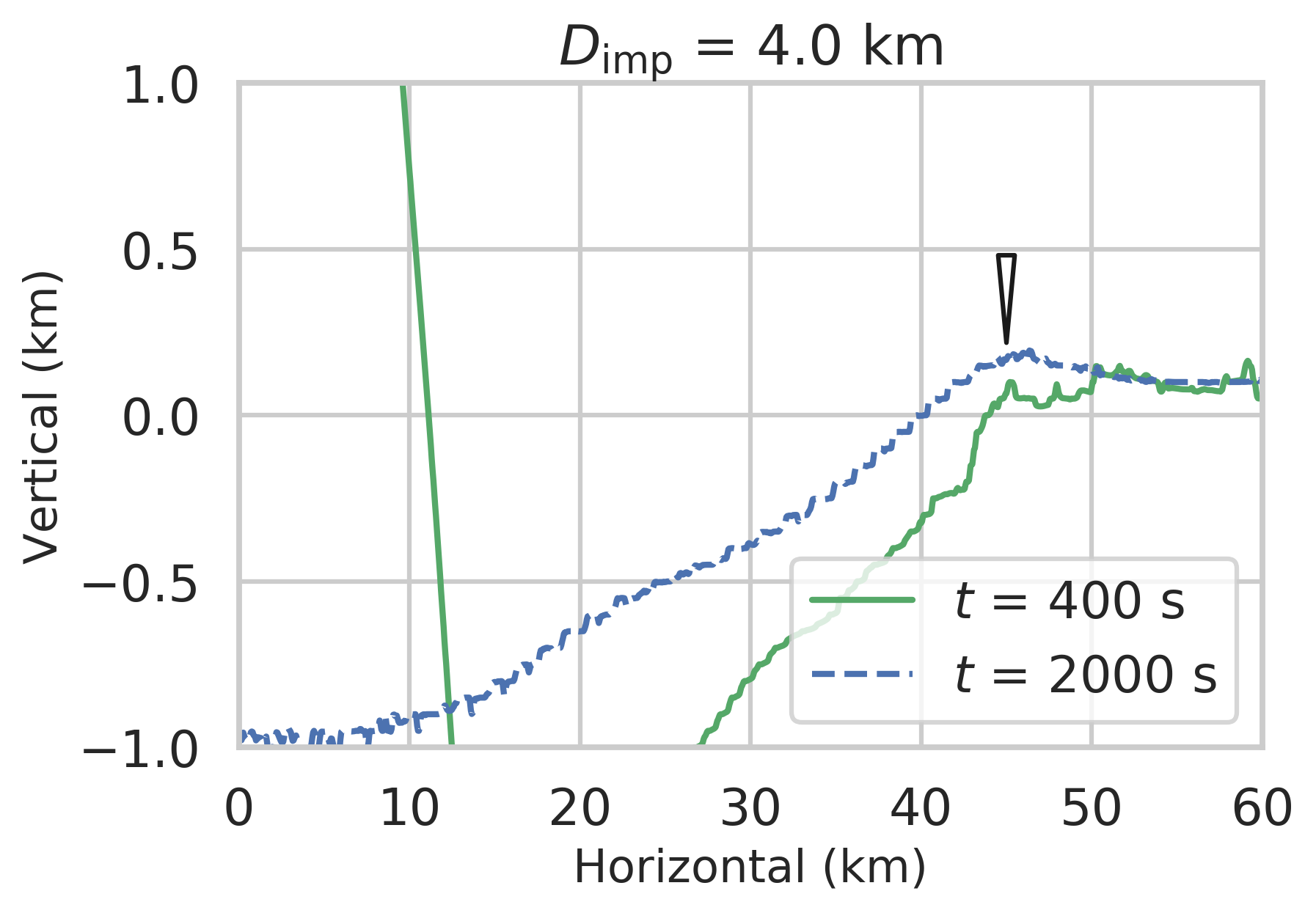}
\caption{
Close-up view of the same impact simulation show in Figure \ref{fig:tmp_15Kb255K}.
Each line shows the surface profile at different time steps; the solid line indicates $t$ = 400 s and the dashed line is $t$ = 2000 s.
Arrows indicate the rim location.
Note the vertical scale is different from Figure \ref{fig:tmp_15Kb255K}.
\label{fig:zoom_15Kb255K}}
\end{figure}

\begin{figure}
\plotone{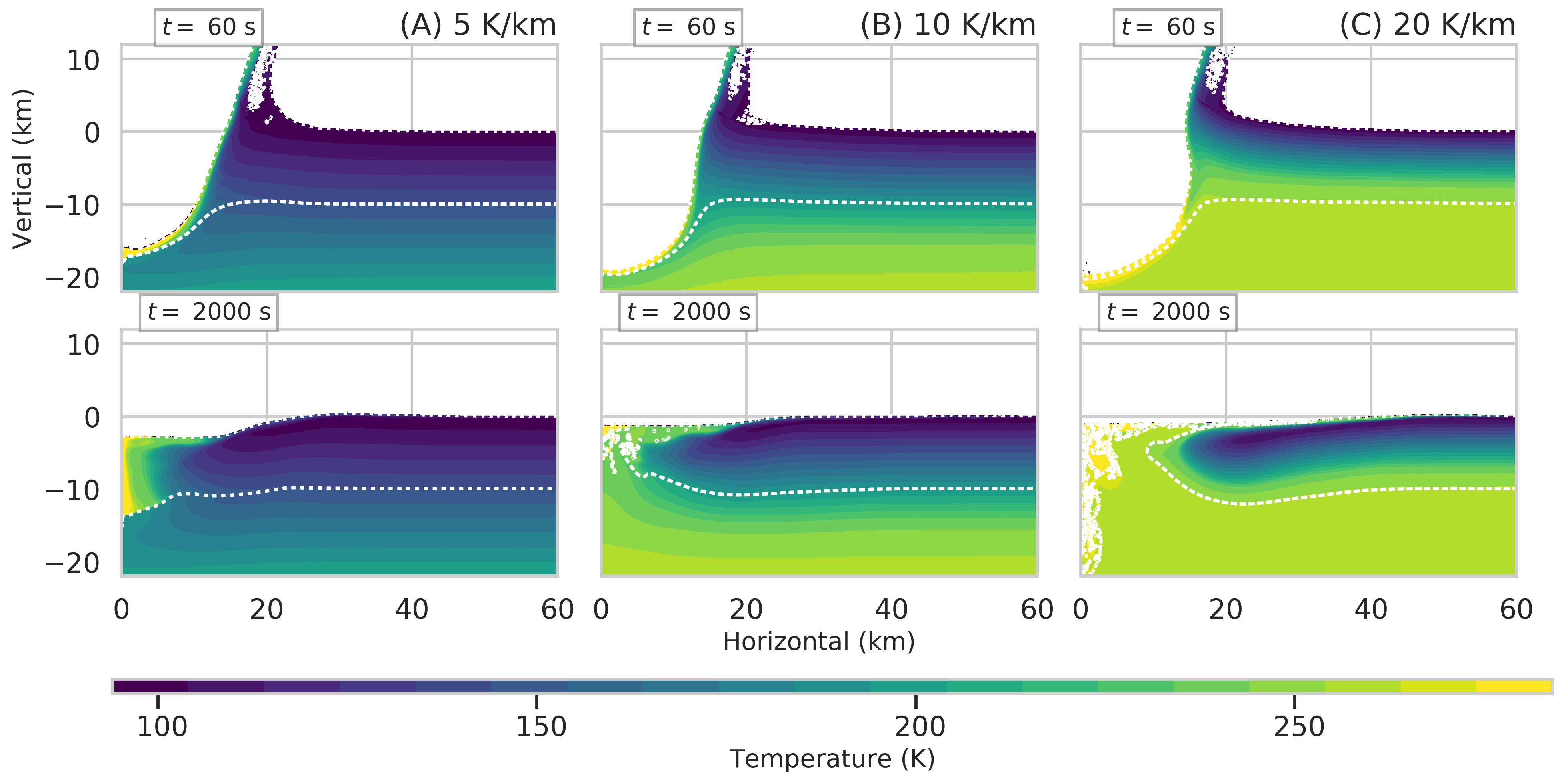}
\caption{
Similar plots to those shown in Figure \ref{fig:tmp_15Kb255K}, but for the different temperature gradient cases; (A) 5 K/km, (B) 10 K/km, and (C) 20 K/km.
Color indicates the temperature in Kelvin, and the white dotted lines represent the material boundary.
\label{fig:tmp_dimp4km_mc10}}
\end{figure}

\appendix
\section{Supplemental Movie} \label{sec:appendix}
\begin{figure}
\begin{interactive}{animation}{MovieA1.mp4}
\plotone{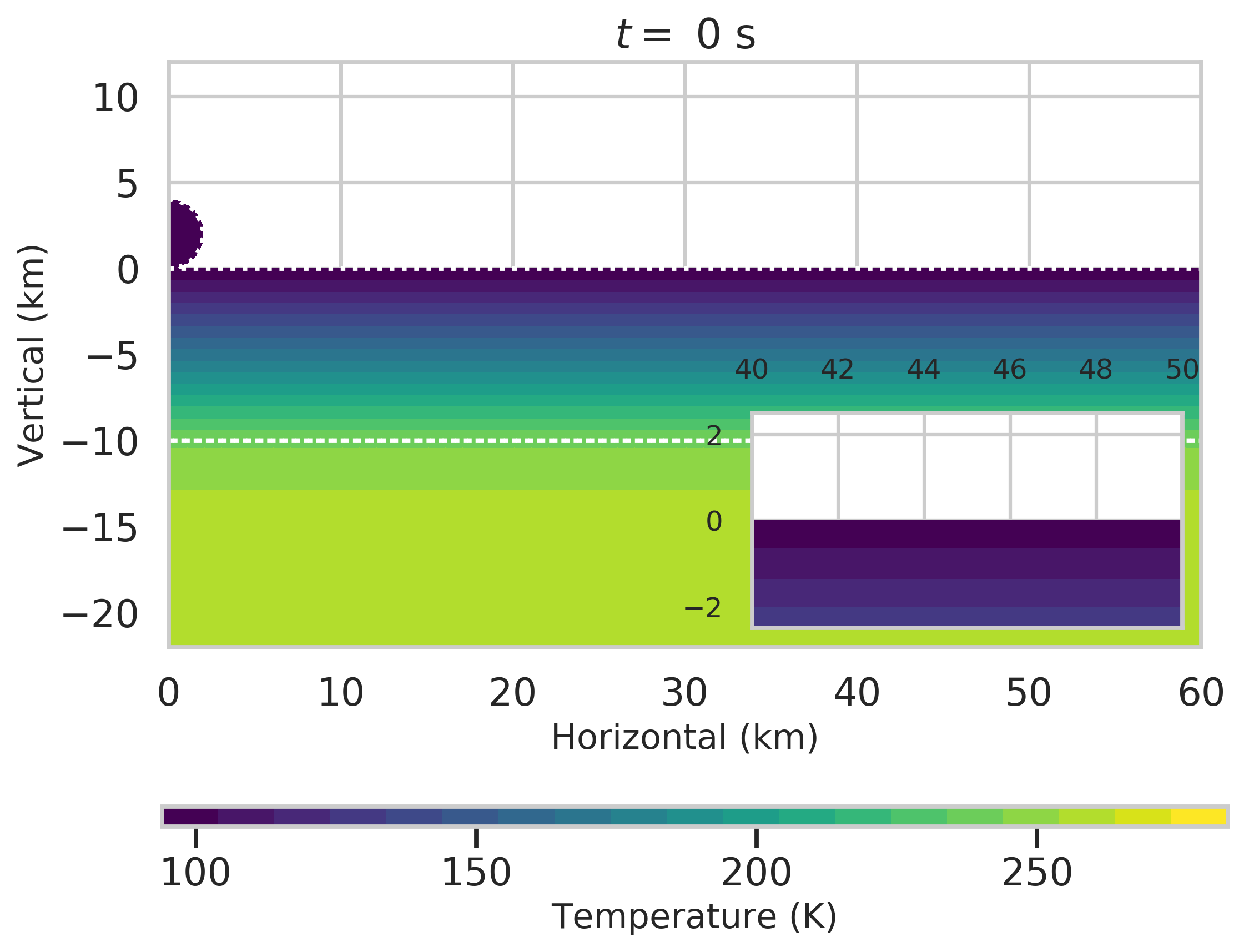}
\end{interactive}
\caption{
Animation for the impact of a 4 km impactor into a 10 km thick methane clathrate layer over water ice with a temperature gradient of 15 K/km.
Additional to snapshots shown in Figure \ref{fig:tmp_15Kb255K}. 
The insert (bottom right) highlights the region around the rim.
The 10 s animation covers 3000 s of the simulation after the impact.
\label{sup:mv}}
\end{figure}

\restartappendixnumbering
Animated Figure \ref{sup:mv} shows impact simulations of a 4 km diameter impactor hitting on 10 km thick methane clathrate with a temperature gradient of 15 K/km, which covers time sequences of the impacts that are missing in Figure \ref{fig:tmp_15Kb255K}.

\end{document}